\documentclass[journal]{IEEEtran}
\ifCLASSINFOpdf
  \usepackage[pdftex]{graphicx}
  \usepackage[table,xcdraw]{xcolor}
\else
\fi
\usepackage{url}

\usepackage{soul}
\usepackage{hyperref}
\usepackage{mathrsfs,amssymb, amsmath}
\begin{document}
%
\title{Context-Aware Convolutional Neural Network for Grading of Colorectal Cancer Histology Images}
%
%

\author{Muhammad~Shaban, Ruqayya~Awan, Muhammad Moazam Fraz,~\IEEEmembership{Senior~Member,~IEEE}, \\Ayesha Azam, David Snead, *Nasir~M.~Rajpoot,~\IEEEmembership{Senior~Member,~IEEE}
\thanks{Manuscript received April 18, 2019. Asterisk indicates corresponding author.}
\thanks{Muhammad Shaban, Ruqayya Awan, Muhammad Moazam Fraz, Ayesha Azam, and Nasir M. Rajpoot are with the Department of Computer Science, University of Warwick, CV4 7AL Coventry, U.K. (e-mail: m.shaban, r.awan.1, moazam.fraz, ayesha.azam, n.m.rajpoot@warwick.ac.uk)}
\thanks{Muhammad Moazam Fraz is also with School of Electrical Engineering and Computer Science, National University of Sciences and Technology, H-12, Islamabad, Pakistan}
\thanks{Ayesha Azam, David Snead, and Nasir~M.~Rajpoot are with the Department of Pathology, University Hospitals Coventry, Warwickshire, UK}
\thanks{Nasir~M.~Rajpoot is also with the The Alan Turing Institute, London, UK}}

\maketitle

\begin{abstract}

Digital histology images are amenable to the application of convolutional neural network (CNN) for analysis due to the sheer size of pixel data present in them. CNNs are generally used for representation learning from small image patches (e.g. $224\times224$) extracted from digital histology images due to computational and memory constraints.
However, this approach does not incorporate high-resolution contextual information in histology images. 
We propose a novel way to incorporate larger context by a context-aware neural network based on images with a dimension of $1,792\times1,792$ pixels. The proposed framework first encodes the local representation of a histology image into high dimensional features then aggregates the features by considering their spatial organization to make a final prediction. The proposed method is evaluated for colorectal cancer grading and breast cancer classification. A comprehensive analysis of some variants of the proposed method is presented. Our method outperformed the traditional patch-based approaches, problem-specific methods, and existing context-based methods quantitatively by a margin of $3.61\%$. Code and dataset related information is available at this link: https://tia-lab.github.io/Context-Aware-CNN
\end{abstract}


\hl{This work has been submitted to the IEEE for possible publication. Copyright may be transferred without notice, after which this version may no longer be accessible.}

%
\IEEEpeerreviewmaketitle

\section{Introduction}
%
%
%
%

\IEEEPARstart{H}istology slides are used by pathologists to analyze the micro-anatomy of cells and tissues through a microscope. However, recent technological developments in digital imaging solutions~\cite{farahani2015whole} have digitized the histology slides (histology images) which enable the pathologists to do the same analysis over the computer screen. These histology images are way larger than natural images, where one cell nucleus usually takes around $50\times50$ square pixels at the highest magnification level (e.g. $40\times$) and each image contains tens of thousands of cells. The digitization process results in an explosion of data which leads to new avenues of research for machine learning and deep learning communities. 

Convolutional neural networks (CNNs) have been widely used to achieve the state-of-the-art results for different histology image analysis tasks such as nuclei detection and classification \cite{koohababni2018nuclei, sirinukunwattana,song2018simultaneous}, metastasis detection \cite{camelyon16, koohbanani2018significance,lin2019fast}, tumor segmentation \cite{qaiser2016persistent} and cancer grading~\cite{arvaniti2018automated,ing2018semantic,RA}. Each task requires a different amount of contextual information, for instance, cell classification needs only high-resolution cell appearance along with little neighboring tissue whereas tumour detection and segmentation rely on a larger context covering multiple cells simultaneously. Due to tumour heterogeneity, cancer grading requires high-resolution cell information as well as the contextual spatial organization of cells in the tumour microenvironment (TME). Most existing CNN based methods applied to histology images follow a patch based approach to training different models tends to ignore contextual information due to memory constraints. Although these models are often trained on a large number of image patches extracted from histology images, often spatial relationships between neighbouring patches are ignored. Due to the lack of large contextual information, the inference is independent of underlying tissue architecture and it is performed based on the limited context captured by individual patches. This approach works well for problems where contextual information is relatively less important for prediction. However, contextual information becomes vital in problems where diagnostic decisions are made on the basis of underlying tissue architecture such as cancer grading.

\begin{figure*}[ht]
   \centering
    \includegraphics[width=0.99\linewidth]{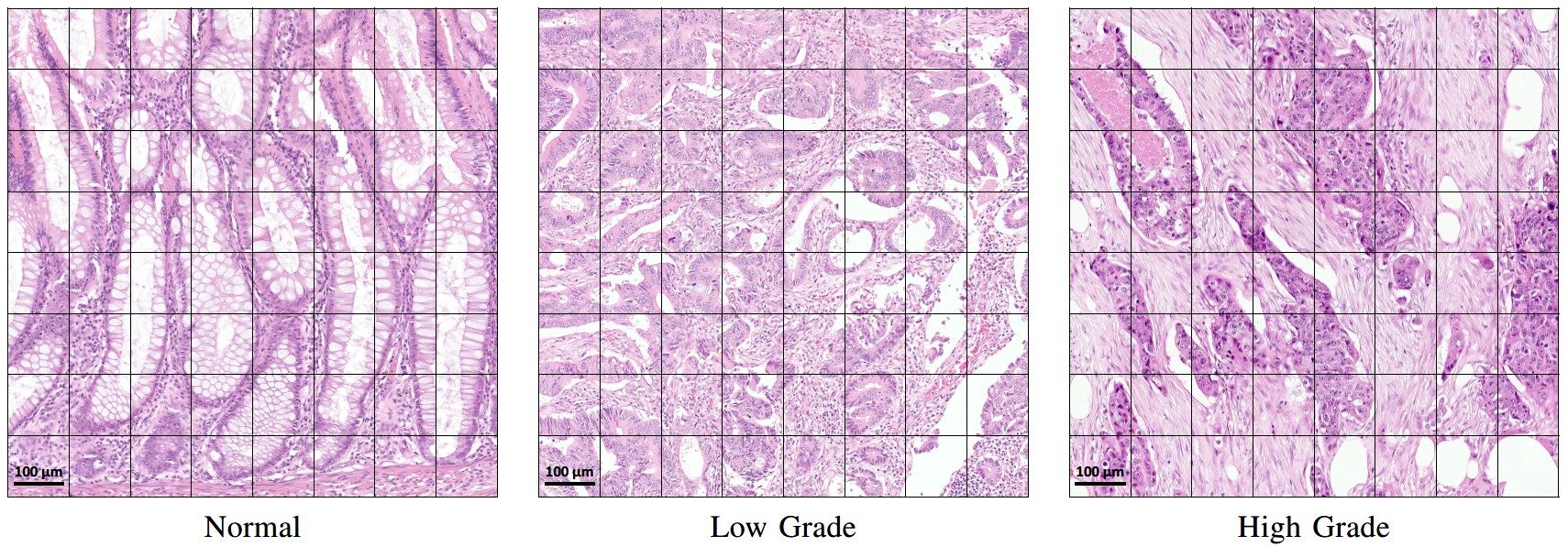}
    \caption{Three visual field regions of colorectal tissue which highlight the importance of larger context for correct grading. Each cell of the overlaid grid shows the $224\times 224$ pixel context captured by a standard patch classifier at $20\times$ magnification.}
    \label{fig_sample}
\end{figure*}

In this paper, we consider colorectal cancer (CRC) grading to demonstrate the significance of context-aware CNNs in cancer histology image analysis. CRC is the fourth most common cause of cancer-related deaths worldwide \cite{wcr}. The grade of CRC is determined by pathologists by collective analysis of individual cancer cells' abnormality and their spatial organization as distorted glandular structure in the histology image. Several studies on the prognostic significance of CRC adopted a two-tiered grading system to reduce the inter-observer variability \cite{binary_grading1, binary_grading2}, merging the well and moderately differentiated glands into a low-grade tumor and classifying tissue with poorly and undifferentiated glands as a high-grade tumor. In this work, we consider diagnostic regions captured from CRC histology images containing enough context to reliably predict the cancer grade (see Figure \ref{fig_sample}). We refer to them as visual fields in this paper as selected by an expert pathologist. A CNN based method for CRC grading requires a high-resolution view of cancer cells in the visual field along with the large contextual information to capture cell organization for accurate grading.

\begin{figure*}[ht]
\centering
\includegraphics[width=6.9in]{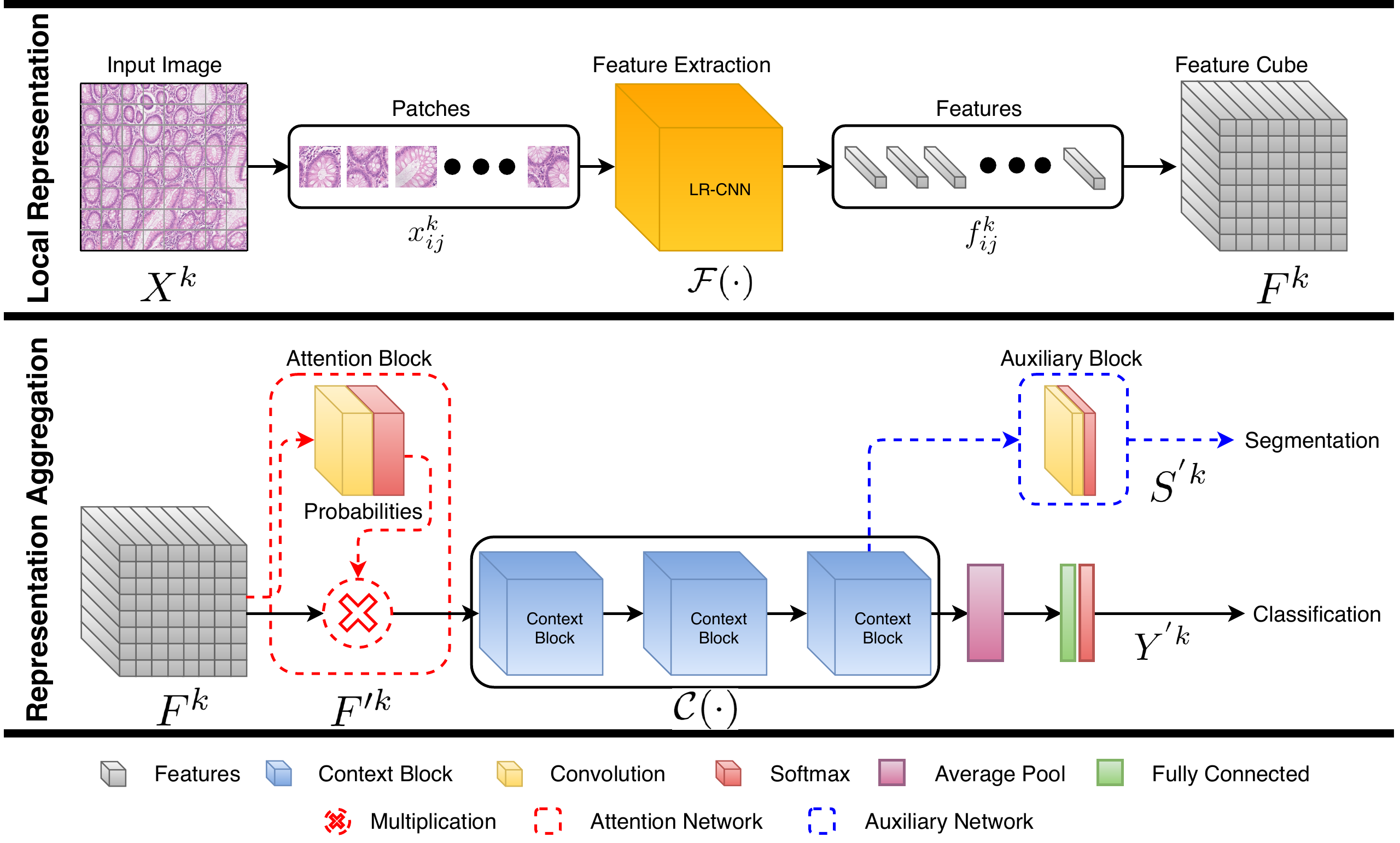}
\caption{Flow diagram of the proposed context-aware framework for CRC grading. The top row shows the local representation learning. The bottom row illustrates the network architecture for representation aggregation learning which consists of multiple context blocks and other standard layers. Dashed lines represent the blocks of a specific network design whereas solid lines represent the common blocks (see Table \ref{tab:notation} for notations).}
\label{fig_flow_diagram}
\end{figure*}

We propose a novel framework for context-aware learning of histology images. The proposed framework first learns the local representation by a CNN (LR-CNN) and then aggregate the contextual information through a representation aggregation CNN (RA-CNN), as shown in Figure~\ref{fig_flow_diagram}. The proposed framework takes $1,792\times 1,792$ input image which is 64 times larger than the standard patch classifier's input. The LR-CNN divides the input image into patches and converts them into high-dimensional feature vectors. These feature vectors are arranged in the form of a feature-cube using the same spatial arrangement in which the corresponding patches were extracted. This feature-cube is then fed into the RA-CNN to make predictions based on both high-resolution feature representation and spatial context. The proposed context-aware framework is flexible enough to incorporate any state-of-the-art image classifier as LR-CNN for local representation learning with the RA-CNN. We present detailed results and that our proposed framework achieves superior performance over traditional patch-based approaches and existing context-aware methods. Moreover, the proposed framework also outperforms the problem specific methods for CRC grading. Our main contributions in this paper are as follows:

\begin{itemize}
\item We propose a novel framework for context-aware learning from large high-resolution input images e.g. $1,792\times1,792$ at $20\times$ resolution.
\item The proposed framework is highly flexible since it can leverage any state-of-the-art network design for local representation learning.
\item We explore different context-aware learning and training strategies to enhance the framework's ability to learn the contextual information.
\item We report comprehensive experiments (with 100+ network models) and comparisons to demonstrate the superiority of the proposed context-aware learning framework over traditional patch-based methods and existing context-aware learning methods.
\end{itemize}

\section{Related Work}
\label{rel_work}
Related work is divided into two subsections: methods related to context-aware learning and some problem specific methods on cancer grading.

\subsection{Context-Aware Learning}
In literature, various different approaches have been presented to incorporate the contextual information for the classification of histology images. Some researchers~\cite{BACH_1, BACH_2, BACH_3} used image down-sampling, a common practice followed in natural image classification, to capture the context from larger histology image. However, this approach is not suitable for problems where cell information is as important as the context. Adaptive patch sampling~\cite{wang2018weakly} and discriminative patch selection~\cite{cruz2018high} from histology images is another way to integrate the sparse context. These methods are not capable of capturing small regions of interest at high resolution e.g. tumor cells and their local contextual arrangement. Some methods~\cite{alsubaie2018multi,bentaieb2017structured,liu2017detecting,koohbanani2018significance} leverage the multi-resolution nature of histology images and use multi-resolution based classifiers to capture context. These multi-resolution approaches only consider a small part of an image at high resolution and the remaining part at low resolutions to make a prediction. Therefore, these approaches lack the contextual information of cellular architecture at high resolution in a histology image.  

Recently, some works~\cite{agarwalla2017representation,kong2017cancer,zanjani2018cancer,li2018cancer} have used larger high-resolution patches to improve the segmentation of histology images. Zanjani \textit{et al.} \cite{zanjani2018cancer} and  Li \textit{et al.}  \cite{li2018cancer} used a CNN based feature extractor followed by a Conditional Random Field (CRF) for context learning. Latter is end-to-end trainable with a patch size of $672\times672$, considerably smaller than the patch size used in our proposed method. Agarawalla \textit{et al.}\cite{agarwalla2017representation} and Kong \textit{et al.}~\cite{kong2017cancer} used a 2D-LSTM instead of CRF to improve tumor segmentation. Some works~\cite{awan2018context,bejnordi2017context} used larger patches at high resolution for the task of context-based classification. Koohbanani \textit{et al.}\cite{awan2018context} proposed a context-aware network for breast cancer classification. They used standard SVM to learn the context from the CNN based features of the patches extracted from a high-resolution image. Due to the nature of the final classifier, this work is only capable of capturing a limited context. Bejnordi \textit{et al.} \cite{bejnordi2017context} proposed a similar approach for breast tissue classification. They trained their network in two steps. In the first step, they used a small patch size and in the second step, they fixed the weights of half of the network to feed a larger patch for training the remaining half of the network. Their network also suffers from a limited context problem as they managed to train a network with the largest patch size of $1,024\times1,024$ pixels with small batch size ($10$ patches). Sirinukunwattana \textit{et al.} \cite{sirinukunwattana2018improving} presented a systematic comparison of different context-aware methods to highlight the importance of context-aware learning.   

As opposed to the aforementioned methods, our proposed method is different in a network design such that it is flexible enough to accommodate any state-of-the-art CNN based image classifiers for representation learning and a custom CNN based architecture for representation aggregation. The representation learning and aggregation are stacked together for context-aware learning with larger image size, $1,792\times1,792$ and a typical batch size of 64 images.

\subsection{Problem Specific Method}
A number of automated methods for objective grading of breast, prostate, and colorectal cancer \cite{naik,farjam, nguyen,RA} have been proposed in the literature. For instance, in \cite{farjam,naik,nguyen}, a linear classifier is trained with handcrafted features based on the glandular morphology for prostate cancer grading. 

Awan \textit{et al.} \cite{RA} presented a method for two-tier CRC grading based on the extent of deviation of the gland from its normal shape (circular/elliptical). They proposed a novel Best Alignment Metric (BAM) for this purpose. As a pre-processing step, CNN based gland segmentation was performed, followed by the calculation of BAM for each gland. For every image, average BAM was considered as a feature along with two more features inspired by BAM values. In the end, an SVM classifier was trained using this feature set for CRC grading.
Our proposed method differs from these existing methods in two ways. First, it does not depend on the intermediate step of gland segmentation making it independent of segmentation inaccuracies. Second, the proposed method is entirely based on a deep neural network which makes this framework independent of cancer type. Therefore, the proposed framework could be used for other context-based histology image analysis problems. In this regard, besides CRC grading, we have demonstrated the application of the proposed method for breast cancer classification.

\section{The Proposed Method}
The proposed framework for context-aware grading consists of two stacked CNNs as shown in Fig~\ref{fig_flow_diagram}. The first network, LR-CNN, converts the high-resolution information of an image into high dimensional feature-cube through patch based feature extraction. The second network, RA-CNN aggregates the learned representation in order to learn the spatial context from the feature-cube to make a prediction. We leverage the power of traditional patch classifiers to learn local representation from individual patches. However, we explore different network architectures for context block in RA-CNN for context-aware learning. Moreover, different training strategies are explored to build a powerful context-aware grading model. The following section explains each building block of the proposed framework in detail. The notation used to describe each building block is summarized in Table \ref{tab:notation}.

\subsection{Network Input}
The input to our framework is an image ($X^k$) from a dataset, $D = \{X^k,Y^k, S^k; k = 1, \dots K\}$, of large high resolution images which consists of $K$ images with corresponding labels $Y^k \in \{1,\dots, C\}$ for classification into $C$ classes and coarse patch level segmentation masks $S^k \in \{1,\dots, C\}$ for multi-task learning. Each image is divided into $M\times N$ patches of same size where $x^{k}_{ij}$ and $y^{k}_{ij}$ represent the ${ij}^{th}$ patch of $k^{th}$ image and its corresponding label, respectively. We used a patch dataset, $d = \{(x^k_{ij},y^k_{ij}), |~x^k_{ij} \in X^k, y^k_{ij} \in Y^k \}$, which consists of patches and their corresponding labels for pre-training of LR-CNN.

\begin{table}[]
\centering
\caption{Enumeration of symbols used in the paper }
\resizebox{8.8cm}{!}{
\begin{tabular}{|c|l|c|l|}
\hline
\textbf{Symbol }            & \textbf{Description}              & \textbf{Symbol} & \textbf{Description}                   \\ \hline
 $D$                          & Image dataset                     & $X^k$         & $k^{th}$ image                \\ \hline
 $K$                          & Number of images                  & $Y^k$         & Label of $k^{th}$ image       \\ \hline
 $C$                      & Number of Classes                 & $S^k$         & Mask of $k^{th}$ image        \\ \hline
$\textbf{X}$                  & Set of all images                   & $\textbf{Y}$    & Labels of \textbf{X}          \\ \hline
$\textbf{S}$                  & Masks of \textbf{X}               & $d$             & Patch dataset                 \\ \hline
 $M$                          & Patches in an image column        & $i$           & $1, \dots, M$                 \\ \hline
 $N$                          & Patches in an image row           & $j$           & $1, \dots, N$                 \\ \hline
 $x^k_{ij}$                 & $ij^{th}$ patch of $X^k$          & $y^k_{ij}$    & Label of $x^k_{ij}$ patch     \\ \hline
 $\mathcal{F}(\cdot )$      & Feature extractor                 & $f^{k}_{ij}$  & Features of $x^k_{ij}$        \\ \hline
 $L_{f}$                    & Fully connected layer             & $L_{p}^{g}$   & Global pooling layer          \\ \hline
 $L_{c}^{a \times a}$       & $a \times a$ convolution layer    & $L_{s}$       & Softmax layer                 \\ \hline
 $\rightarrow$              & Transition between layers         & $\bullet$     & Preceding layer's output      \\ \hline
 $\otimes$                  & Hadamard product                  & $\oplus$      & Feature Concatenation         \\ \hline
$\mathcal{B}(\cdot)$        & Context-block                     & $\mathcal{C}(\cdot)$    & Context-Net         \\ \hline
$\textbf{F}$                 & Feature of \textbf{X}             & $\textbf{F}'$ & Weighted Feature of \textbf{X}\\ \hline
 $\textbf{Y}'$              & Predicted labels of \textbf{X}    & $Y'^k$        & Predicted label of $X^k$     \\ \hline
 $\textbf{S}'$              & Predicted Masks of \textbf{X}     & $S'^k$        & Predicted Mask of $X^k$     \\ \hline
 $W^k$      &  $k^{th}$ image weight              & $\theta$       &  Learnable Parameters           \\ \hline
 $\mathcal{L}_{cls}$        & Classification cost function      & $\mathcal{L}_{wgt}$ &  Weighted cost function           \\ \hline
 $\mathcal{L}_{seg}$ &  Segmentation cost function              & $\mathcal{L}_{joint}$      &  Joint cost function           \\ \hline
\end{tabular}}
\label{tab:notation}
\end{table}

\subsection{Local Representation Learning}
\label{sec:MFE}
First part of the proposed framework encodes an input image $X^k$ into a feature-cube $F^k$. All the input images are processed through the LR-CNN in a patch based manner. The proposed framework is flexible enough to use any state-of-the-art image classifier as a LR-CNN such as ResNet50~\cite{resnet}, MobileNet~\cite{mobilenet}, Inception~\cite{inception}, or Xception~\cite{xception}. This flexibility also enables it to use pre-trained weights in case of a limited dataset. Moreover, it is possible to train the LR-CNN independently before plugging it into the proposed framework, enabling it to learn meaningful representation \cite{zhang2017mdnet} which leads to early convergence of the context-aware learning part of the framework.

\subsection{Feature Pooling}
\label{sec:MFP}
The spatial dimensions of the output feature $f^k_{ij}$ of a patch $x^k_{ij}$ may vary depending on the input patch dimensions and the network architecture for feature extraction. A global feature pooling layer is employed to get a similar dimensional feature vector for all variations of the proposed framework. Both average and max global pooling strategies are explored. After global pooling, features of all patches are rearranged in the same spatial order ($M\times N$) as extracted patches to construct the feature-cube $F^k$ for context-aware learning. The depth of the feature-cube depends on the choice of LR-CNN. For the sake of generality, we will represent the output of our LR-CNN as follows,

\begin{equation}
\label{eq:cnn}
\textbf{F} = \mathcal{F}(\textbf{X},\theta_{\mathcal{F}}) \rightarrow L_{p}^{g}(\bullet)
\end{equation} where $\mathcal{F}$ represents the fully convolutional part of the LR-CNN and acts as a feature extractor whereas \textbf{X} is the batch of images and \textbf{F} is the local feature representation of \textbf{X} after pooling $L_p^{g}$, which could be a global average or max pooling layer. The operator ($\rightarrow$) provides the output of the preceding layer to the following layer and operator ($\bullet$) represents the output of the preceding layer. 

\subsection{Feature Attention}
As the input to the proposed framework has a relatively large spatial dimension, there may be some part of the image that may not have any significance for the prediction of image label. We introduce an attention block to give less weight to insignificant features and vice-versa. This attention block takes feature-cube as input and learns the weight of usefulness for each value in the feature-cube. Hadamard product is taken between the weights and input feature-cube to increase the impact of more important areas of an image in label prediction and vice-versa. The weighted feature-cube $\textbf{F}'$ is defined as: 

\begin{equation}
\label{eq:attn}
\textbf{F}' = L_{c}^{1\times 1}(\textbf{F},\theta_{c}) \rightarrow L_s(\bullet) \otimes \textbf{F},
\end{equation} where $L_{c}^{1\times 1}$ and $\theta_{c}$ represent the $1\times  1$ convolution layer and its parameters, respectively. $L_s$ denotes the softmax layer and the operator $\otimes$ is used to represent Hadamard product.

\subsection{Context Blocks}
\label{sec:MCB}
Since the LR-CNN is used to encode the important patch-based image representation into a feature-cube, therefore the main aim of the context block (CB) is to learn the spatial context within the feature cube. The CB learns the relation between the features of the image patches considering their spatial location. We propose three different CB architectures, each with different complexity and capability to capture the context information. First CB, $\mathcal{B}_1(\cdot)$, is comprised of a $3\times 3$ convolution layer followed by ReLU activation and batch normalization. 
Second CB, $\mathcal{B}_2(\cdot)$, uses residual block \cite{resnet} architecture with two different filter sizes. It consists of three convolution layers each followed by batch normalization and ReLU activation. The first and last layers are with $1\times 1$ convolution filter to squeeze and expand the feature depth. The output feature-maps of the last layer are concatenated with the input features-maps which makes its final output. The $\mathcal{B}_2(\cdot)$ is defined as:

\begin{equation}
\label{eq:cb_2}\begin{aligned}
\mathcal{B}_2(\textbf{F}',\theta_{\mathcal{B}_2}) & = [L_{c}^{1\times 1}(\textbf{F}',\theta_{\mathcal{B}^1_2}) \rightarrow L_{c}^{3\times 3}(\bullet,\theta_{\mathcal{B}^2_2}) \\ 
&\rightarrow L_{c}^{1\times 1}(\bullet,\theta_{\mathcal{B}^3_2})] \oplus \textbf{F}',
\end{aligned}\end{equation} where $L_{c}^{1\times 1}$ and $L_{c}^{3\times 3}$ denote the convolution layers with $1\times 1$ and $3\times 3$ filter sizes; $\theta_{\mathcal{B}^1_2}$, $\theta_{\mathcal{B}^2_2}$, and $\theta_{\mathcal{B}^3_2}$ are the parameters of different convolution layers and $\theta_{\mathcal{B}_2}$ represents parameter of the whole context block for brevity. The operator $\oplus$ represents the concatenation of feature-maps.

Unlike the previous two context blocks, our third CB processes the input feature-maps in parallel with different filter sizes to capture context from varying receptive fields. Similar to the blocks in \cite{inception}, it consists of multiple $1\times 1$ and $3\times 3$ convolution layers each followed by batch normalization and ReLU activation. A $3\times 3$ average pooling layer $L_{p}^{3\times 3}$ is also used to average the local context information. The CB, $\mathcal{B}_3$, is defined as:
\begin{equation}
\label{eq:cb_3}\begin{aligned}
\mathcal{B}_3(\textbf{F}',\theta_{\mathcal{B}_3}) & = [L_{c}^{1\times 1}(\textbf{F}',\theta_{\mathcal{B}^1_3}) \rightarrow L_{c}^{3\times 3}(\bullet,\theta_{\mathcal{B}^2_3}) \\ 
& \rightarrow L_{c}^{3\times 3}(\bullet,\theta_{\mathcal{B}^3_3})] \oplus [L_{c}^{1\times 1}(\textbf{F}',\theta_{\mathcal{B}^4_3})] \\
& \oplus [L_{c}^{1\times 1}(\textbf{F}',\theta_{\mathcal{B}^5_3}) \rightarrow L_{c}^{3\times 3}(\bullet,\theta_{\mathcal{B}^6_3})] \\
& \oplus [L_{p}^{3\times 3}(\textbf{F}') \rightarrow L_{c}^{1\times 1}(\bullet,\theta_{\mathcal{B}^7_3})],
\end{aligned}\end{equation} where $\theta_{\mathcal{B}^1_3}$ to $\theta_{\mathcal{B}^7_3}$ are the parameters of different convolution layers and  $\theta_{\mathcal{B}_3}$ represents parameter of the whole context block for the sake of notational simplicity.


\subsection{Representation Aggregation for Context Learning}
\label{sec:MCAN}
The local representation tissue has been learned by the LR-CNN. Therefore, the task of spatial context learning from feature-cube is relatively less challenging as compared to context learning from the raw image. A cascaded set of three context blocks ($\mathcal{C}(\cdot )$) of the same type ($\mathcal{B}_1$,$\mathcal{B}_2$, or $\mathcal{B}_3$) is used in RA-CNN. These context blocks are explained in section \ref{sec:MCB}. The output of $\mathcal{C}(\cdot )$ is followed by a global average pooling, a fully connected, and a softmax layer to make the final prediction in the required number of classes. The final prediction $\textbf{Y}'$ from the features of input images $\textbf{X}$ is computed as:
 
\begin{equation}
\label{eq:can}
\textbf{Y}' = \mathcal{C}(\textbf{F}',\theta_{\mathcal{C}}) \rightarrow L_{p}^{g}(\bullet) \rightarrow L_{f}(\bullet,\theta_{f'}) \rightarrow L_{s}(\bullet),
\end{equation} where $\theta_{\mathcal{C}}$ and $\theta_{f'}$ represent the parameters of all context blocks and the fully connected layer in RA-CNN, respectively. The proposed framework is trained end-to-end with categorical cross-entropy loss based cost function $\mathcal{L}_{cls}(\cdot)$ which is defined as:

\begin{equation}
\label{eq:ce_cls}
\mathcal{L}_{cls}(\textbf{Y},\textbf{Y}') = -\frac{1}{K} \sum_{k=1}^{K} \sum_{c=1}^C {Y^k_c\log_{2}(Y^{'k}_c)},
\end{equation} where $Y^k_c$ and $Y^{'k}_c$ are the ground truth and predicted probabilities of $k^{th}$ image for $c^{th}$ class.

\subsection{Auxiliary Block}
\label{sec:MAN}
The proposed framework is designed for the classification of large input images. Therefore, the label of an input image may depend on a set of different primitive structures (such as glands) and their spatial organization. To exploit these primitive structures for better classification we proposed an auxiliary block which acts as patch based segmentation of the primitive structures in the input image. This will improve the convergence of proposed networks and also output the coarse patch based segmentation mask ($S_s'$) along with image label ($Y'$). The segmentation masks ($\textbf{S}'$) of input images $\textbf{X}$ from their features $\textbf{F}'$ is defined as:

\begin{equation}
\label{eq:seg}
\textbf{S}' = \mathcal{C}(\textbf{F}',\theta_{\mathcal{C}}) \rightarrow  L_{c}^{1\times 1}(\bullet,\theta_{c'}) \rightarrow L_{s}(\bullet),
\end{equation} where $L_{c}^{1\times 1}$ is a convolution layer with $\theta_{c'}$ parameters. 
The addition of auxiliary block enables the proposes framework to learn in a multi-task setting, where the coarse segmentation-map guides the network to improve the individual patch based feature classification in addition to the prediction of the input image. This leads to a network with improved classification performance since it is minimizing both segmentation and classification loss simultaneously. The segmentation-map based loss function ($\mathcal{L}_{seg}$) and joint loss function ($\mathcal{L}_{joint}$) are defined as:

\begin{equation}
\label{eq:ce_seg}
\mathcal{L}_{seg}(\textbf{S},\textbf{S}') = -\frac{1}{K} \sum_{k=1}^{K} \sum_{c=1}^C {S^k_c\log_{2}(S^{'k}_c)},
\end{equation}

\begin{equation}
\label{eq:joint}\begin{aligned}
\mathcal{L}_{joint}(\textbf{Y},\textbf{Y}',\textbf{S},\textbf{S}') = & \alpha \times \mathcal{L}_{cls}(\textbf{Y},\textbf{Y}') + \\
&(1 - \alpha )  \times \mathcal{L}_{seg}(\textbf{S},\textbf{S}'),
\end{aligned}\end{equation} where $\alpha$ is a hyper-parameter which defines the contribution of both loss functions in the final loss. Similar to patch classifier, the loss function ($\mathcal{L}_{joint}$) is minimized with RMSprop optimizer~\cite{rmsprob}.

\subsection{Training Strategies}
\label{sec:train}
We trained the proposed framework in four different ways with varying ability to capture the spatial context. First, the proposed framework is trained without attention block and by minimizing the $\mathcal{L}_{cls}(\cdot)$ loss only. This configuration is represented by solid line blocks in Fig~\ref{fig_flow_diagram}. Second, the same configuration as first but trained with a sample-based weighted loss function, $\mathcal{L}_{wgt}(\cdot)$, which give more weight to the image patches with relatively less region of interest (glandular region) as compared to the background. The weight of an image and $\mathcal{L}_{wgt}(\cdot)$ are defined as follow,
\begin{equation}
\label{eq_weighted}
W^k = 
\begin{cases}
  \frac{1}{R^k_{roi}}, & \text{if } R^k_{roi} > \alpha\\
  \frac{1}{\alpha}, & \text{otherwise}
\end{cases}
\end{equation}
\begin{equation}
\label{eq_weighted_loss}
\mathcal{L}_{wgt}(\textbf{Y},\textbf{Y}') = -\frac{1}{K} \sum_{k=1}^{K} \sum_{c=1}^C {W^k Y^k_c\log_{2}(Y^{'k}_c)},
\end{equation} where $R^k_{roi}$ and $W^k$ represent the ratio of the region of interest and weight of the $k^{th}$ image. The $\alpha$ is the ratio threshold, selected empirically as 0.10, sets the upper limit of an image weight. Third, multi-task learning based training with the help of an auxiliary block by using joint classification and segmentation loss, $\mathcal{L}_{joint}$. Last, training using the same joint loss but with attention-based feature-cube to amplify the contribution of more important features in the feature-cube. The network configuration of this strategy is represented by both solid and dotted lines blocks in Fig~\ref{fig_flow_diagram}. We termed these strategies as \textit{standard}, \textit{weighted}, \textit{auxiliary}, and \textit{attention}, respectively.


\section{Datasets \& Performance Measures}
In this section, we explain the dataset details used for training and evaluation of the proposed framework and metrics for performance evaluation.

\subsection{Datasets}
The proposed framework is evaluated on two different tissue types for two different tasks in order to demonstrate its capabilities. Our colorectal cancer dataset \cite{RA} is comprised of visual fields (refer as images for simplicity) extracted from colorectal histology images based on a two-tier \cite{binary_grading1, binary_grading2} grading system. The CRC dataset consists of 139 images with an average size of $4,548\times 7,520$ pixels obtained at $20\times$ magnification. These images are classified into three different classes (normal, low grade, and high grade) based on the organization of glands in the images by the expert pathologist. We follow 3-fold cross validation for a fair comparison of the proposed method with the method presented in \cite{RA}. 
We extracted patches of two different sizes for the training of traditional patch classifiers ($224\times224$ pixels) and our proposed framework ($1,792\times1,792$ pixels). A detailed distribution of the patches is presented in table \ref{tab:dataset}. We introduced background class to handle the patches with no or little glandular regions.
In breast cancer classification~\cite{araujo2017classification}, our goal is to classify the images into four different tissue sub-types namely normal, benign, in-situ, and invasive. We use exactly the same training, validation, and test splits as in \cite{sirinukunwattana2018improving} for a fair comparison of our proposed framework with the existing context-aware methods.

\begin{table}[]
\centering
\caption{Number of patches in each class and fold of training dataset.}
\begin{tabular}{|c|c|c|c|c|}
\hline
\multicolumn{5}{|c|}{Patches of size 224$\times$224}        \\ \hline
Folds & Background & Normal & Low Grade & High Grade \\ \hline
1     & 25,000      & 25,000  & 25,000     & 25,000      \\ \hline
2     & 25,000      & 25,000  & 25,000     & 25,000      \\ \hline
3     & 25,000      & 25,000  & 25,000     & 25,000      \\ \hline
Total & 75,000      & 75,000  & 75,000     & 75,000      \\ \hline
\multicolumn{5}{|c|}{Patches of size 1,792$\times$1,792}      \\ \hline
1     & 1,750       & 3,500   & 3,500      & 3,500       \\ \hline
2     & 1,750       & 3,500   & 3,500      & 3,500       \\ \hline
3     & 1,750       & 3,500   & 3,500      & 3,500       \\ \hline
Total & 5,250       & 10,500  & 10,500     & 10,500      \\ \hline
\end{tabular}
\label{tab:dataset}
\end{table}

\subsection{Performance Measures} 
We have used three metrics, the average accuracy, F1 score and Rank-sum measure for performance evaluation. The average accuracy refers to the percentage of images classified correctly, across the three folds. 
Rank-sum based evaluation metric is used to summarize the accuracy of different models trained using a specific setting in order to compare models trained with different context-blocks and LR-CNNs. Different colors are used to represent different rank for better illustrative visualization as shown in Table \ref{tab:CAN}, \ref{tab:Pool} and \ref{tab:comp2}. The orange color indicates the best performing method whereas green, blue, yellow, and red colours indicate that the results are within 97.5\%, 95\%, 90\%, and 85\% of the best performing method, respectively. The rank for these colors are: orange = 1, green = 2, blue = 3, yellow = 4, red = 5, and no colour = 6. The lowest rank-sum shows the best performance.

\section{Experiments \& Results}
The proposed framework is extensively evaluated and compared with existing approaches in three different categories, i.e. traditional patch based classifiers, existing context-aware approaches and domain oriented methods for CRC grading. The details of experimental evaluation are given in following subsections.

\subsection{Experimental Setup}
The CRC images are divided into patches of size $1,792\times 1,792$, and the label of each patch is predicted using the proposed framework with a stride of $224\times 224$. To avoid redundant processing of the same region, the input images are processed with LR-CNN to get representation features of each local region. Afterwards, RA-CNN is applied in a sliding window manner to aggregate local representation for context-aware predictions. Through this approach, we process the input image with a 64 times bigger context as compared to standard patch classifier, with only 10\% additional processing time.  The majority voting among predicted grades (normal, low grade and high grade) is used to obtain the final grade of an image. 
The performance of the different variants of the proposed framework is evaluated for the task of CRC grading to show its stability and superior performance over other methods. These variations include the use of four different state-of-the-art classifiers for local representation learning in LR-CNN; spatial dimensionality reduction through average and max-pooling; the usage of three different context-blocks in RA-CNN; and four different training strategies. By employing different combinations of above-mentioned variations, we trained around 100 models in total for each fold of CRC training dataset. Note that, all reported results for CRC grading are image based not patch based. The details of these combinations and their results are given in the following subsection.

\subsection{LR-CNN based Classifiers}
Four different LR-CNNs are trained using (ResNet50~\cite{resnet}, Inception~\cite{inception}, MobileNet~\cite{mobilenet}, and Xception~\cite{xception}) with patch size of $224\times 224$ to get the baseline CRC grading results.
The ResNet-50~\cite{resnet} and Inception network are the winner of Image-Net~\cite{alexnet} challenge in 2015 and 2016, respectively. MobileNet is a lightweight network with just 3 million parameters whereas Xception network uses separable convolutions which results in a significant reduction in computational complexity. The performance of these classifiers for CRC grading is reported in Table \ref{tab:FE}. Although, the performance of all classifiers is comparable, MobileNet shows superior performance with highest mean accuracy. On the other hand, Xception classifier shows consistent performance across three folds with the lowest standard deviation (Std.). 

\begin{table}[ht]
\centering
\caption{Accuracy comparison of four patch classifiers.}
\begin{tabular}{|l|c|c|c|c|c|}
\hline
\rowcolor[HTML]{EFEFEF} 
\multicolumn{1}{|c|}{Network} & Fold-1 & Fold-2  & Fold-3  & Mean    & Std.   \\ \hline
ResNet50                    & 93.48 & 93.62 & 89.13 & 92.08 & 2.08 \\ \hline
MobileNet                    & 93.48 & \textbf{95.74} & 89.13 & \textbf{92.78} & 2.74 \\ \hline
Inception-v3                  & \textbf{95.65} & 91.49 & 86.96 & 91.37 & 3.55 \\ \hline
Xception                    & 93.48 & 91.49 & \textbf{91.30} & 92.09 & \textbf{0.98} \\ \hline
\end{tabular}
\label{tab:FE}
\end{table}

\subsection{RA-CNN based Context-Aware Learning}
We experimented with three context-blocks, $\mathcal{B}_1$, $\mathcal{B}_2$, and $\mathcal{B}_3$, to train three different variation of RA-CNN, which we termed as RA-CNN 1, RA-CNN 2, and RA-CNN 3. These three RA-CNN classifiers are trained separately with all four LR-CNNs as explained in section \ref{sec:MCAN}, hence giving 12 different combinations of the context-aware network. The rank-sum method is used to compare the CRC grading performance of these networks with each other and also with the LR-CNNs. The results in table \ref{tab:CAN}, shows that context-aware networks achieve superior performance as compare to standard patch based classifiers (LR-CNNs). The RA-CNN 3 achieve the best Rank-sum (lowest) which shows its robustness across different representation learning networks. Other two context-aware networks also show comparable performance by remaining in the 97.5\% of the best performer.

\begin{table}[ht]
\centering
\caption{Accuracy comparison of three different context-aware networks with standard patch classifiers.}
\resizebox{8.8cm}{!}{
\begin{tabular}{|l|c|c|c|c|}
\hline
\rowcolor[HTML]{EFEFEF} 
\multicolumn{1}{|c|}{LR-CNN (Avg)} & Baseline                                   & RA-CNN 1                                & RA-CNN 2                                & RA-CNN 3                                \\ \hline
ResNet50                & \cellcolor[HTML]{9AFF99}92.08$\pm$2.08 & \cellcolor[HTML]{FF9F33}94.25$\pm$2.70                & \cellcolor[HTML]{9AFF99}92.08$\pm$2.08 & \cellcolor[HTML]{9AFF99}93.51$\pm$3.10 \\ \hline
MobileNet               & \cellcolor[HTML]{9AFF99}92.78$\pm$2.74 & \cellcolor[HTML]{9AFF99}93.52$\pm$3.55 & \cellcolor[HTML]{9AFF99}93.52$\pm$1.78 & \cellcolor[HTML]{FF9F33}94.25$\pm$2.70                \\ \hline
InceptionV3             & \cellcolor[HTML]{33ACFF}91.37$\pm$3.55 & \cellcolor[HTML]{9AFF99}94.23$\pm$3.71 & \cellcolor[HTML]{9AFF99}94.96$\pm$2.72 & \cellcolor[HTML]{FF9F33}95.68$\pm$1.78                \\ \hline
Xception                & \cellcolor[HTML]{33ACFF}92.09$\pm$0.98 & \cellcolor[HTML]{9AFF99}94.96$\pm$2.72 & \cellcolor[HTML]{9AFF99}94.96$\pm$2.72 & \cellcolor[HTML]{FF9F33}95.68$\pm$3.55                \\ \hline
Rank-sum                    & 10                                   & 7                                    & 8                                    & \textbf{5}                                    \\ \hline
\end{tabular}}
\label{tab:CAN}
\end{table}

\subsection{Local Representation Robustness}
We also conducted different experiments to analyze the robustness of local representation learned by different LR-CNNs. These LR-CNNs are used in combination with different RA-CNNs for context learning along with different feature pooling strategies. Each LR-CNN is used to training three RA-CNNs with both global average and global max pooled feature-cubes. The table \ref{tab:Pool} compares the results using Rank-sum method. It can be observed that the Xception model turns-out as most robust feature extractor in LR-CNNs with the best rank-sum score of 8. Inception model shows comparable results to the best performer as its network design has significant overlap with Xception architecture. 
\begin{table}[]
\centering
\caption{Robustness analysis of feature extractors across different methods.}
\resizebox{8.8cm}{!}{
\begin{tabular}{|l|c|c|c|c|}
\hline
\rowcolor[HTML]{EFEFEF}
\multicolumn{1}{|c|}{Methods}   & ResNet50 (\%)                        & MobileNet(\%)                        & InceptionV3(\%)                      & Xception(\%)                         \\ \hline
RA-CNN 1 (Avg)                      & \cellcolor[HTML]{9AFF99}94.25$\pm$2.70 & \cellcolor[HTML]{9AFF99}93.52$\pm$3.55 & \cellcolor[HTML]{9AFF99}94.23$\pm$3.71 & \cellcolor[HTML]{FF9F33}94.96$\pm$2.72                \\ \hline
RA-CNN 1 (Max)                      & \cellcolor[HTML]{9AFF99}93.52$\pm$1.87 & \cellcolor[HTML]{9AFF99}93.51$\pm$3.10 & \cellcolor[HTML]{FF9F33}94.23$\pm$2.07                & \cellcolor[HTML]{9AFF99}93.54$\pm$3.03 \\ \hline
RA-CNN 2 (Avg)                      & \cellcolor[HTML]{33ACFF}92.08$\pm$2.08 & \cellcolor[HTML]{9AFF99}93.52$\pm$1.78 &\cellcolor[HTML]{FF9F33}94.96$\pm$2.72                & \cellcolor[HTML]{FF9F33}94.96$\pm$2.72                \\ \hline
RA-CNN 2 (Max)                      & \cellcolor[HTML]{FF9F33}95.68$\pm$3.55                & \cellcolor[HTML]{9AFF99}93.52$\pm$3.55 & \cellcolor[HTML]{33ACFF}92.80$\pm$2.72 & \cellcolor[HTML]{9AFF99}93.54$\pm$3.03 \\ \hline
RA-CNN 3 (Avg)                      & \cellcolor[HTML]{9AFF99}93.51$\pm$3.10 & \cellcolor[HTML]{9AFF99}94.25$\pm$2.70 & \cellcolor[HTML]{FF9F33}95.68$\pm$1.78                & \cellcolor[HTML]{FF9F33}95.68$\pm$3.55                \\ \hline
RA-CNN 3 (Max)                      & \cellcolor[HTML]{9AFF99}94.23$\pm$2.07 & \cellcolor[HTML]{9AFF99}92.82$\pm$2.01 & \cellcolor[HTML]{9AFF99}94.25$\pm$2.70 & \cellcolor[HTML]{FF9F33}94.96$\pm$2.72                \\ \hline
Rank-sum & 12                                   & 12                                   & 10                                   & \textbf{8}                                    \\ \hline
\end{tabular}}
\label{tab:Pool}
\end{table}

\subsection{Training Strategies}
\begin{figure*}[ht]
   \centering
    \includegraphics[height=0.45\linewidth]{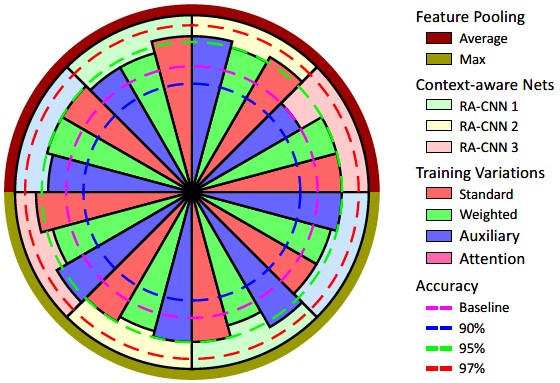}                 
    \caption{(Left) Results of 24 experiments using best performing local representation features (Xception). (Right) Legend  represents the feature pooling type, context-aware net and training strategies used for the experiments. Baseline accuracy is the accuracy of standard patch based Xception classifier.}
    \label{fig:all_xception}
\end{figure*}

We experimented with four different context related training strategies (\textit{Standard}, \textit{Weighted}, \textit{Auxiliary} and \textit{Attention}) to explored their impact on overall performance. The details of each training strategy are given in Section \ref{sec:train}. Table \ref{tab:TS} shows the comparison of these training strategies for Xception based LR-CNN. Each entry in the table contains the average accuracy across three RA-CNNs for particular feature pooling (shown in rows) and the training strategies (in columns). Attention-based training shows the superior results for max-pooled features whereas standard training strategy achieves comparable performance for average-pooled features. However, auxiliary loss based training remains robust for both pooling types and achieves the best overall accuracy. More importantly, each model shows superior performance than the baseline LR-CNN classifier as shown in Fig. \ref{fig:all_xception}. The graphical illustration of 24 experiments using the best performing LR-CNN is shown in Fig. \ref{fig:all_xception}. The results obtained with different combinations of feature pooling type, the context blocks in RA-CNN and the training strategies used for the experiments are illustrated in the pie-chart format for better visual comparison. The accuracy obtained by Xception based LR-CNN is considered as the baseline for comparative analysis. Pie-charts for results with other LR-CNNs are given in supplementary material.

\begin{table}[]
\centering
\caption{Comparison for different training strategies with Xception as feature extractor.}
\begin{tabular}{|l|c|c|c|c|}
\hline
\rowcolor[HTML]{EFEFEF} 
\multicolumn{1}{|c|}{Feature} & Standard       & Weighted & Auxiliary      & Attention      \\ \hline
Xception - Max                & 94.01          & 94.49    & 94.73          & \textbf{95.21} \\ \hline
Xception - Avg                & \textbf{95.20} & 94.72    & 94.72          & 94.00          \\ \hline
Mean                          & 94.61          & 94.60    & \textbf{94.72} & 94.61          \\ \hline
\end{tabular}
\label{tab:TS}
\end{table}

\subsection{Comparative Evaluation}
The proposed method is compared with both domain-based and other context-based methods. Awan \textit{et al.}~\cite{RA} presented a two-step problem specific method for CRC grading where the first step is deep learning based gland segmentation and in the second step the domain knowledge is exploited by computing the Best Alignment Metric (BAM) to capture the shape difference of CRC glands from their expected shape (shape of normal gland). They experimented with two different feature sets which we refer as BAM-1 and BAM-2 in this paper. BAM-1 comprises of average BAM and BAM entropy while BAM-2 comprises of an additional feature known as regularity index. Their method achieved good accuracy for binary classification, normal vs cancer, however, it lakes the robustness required for multi-class grading of CRC images (see Table \ref{tab:comp1}). Our best performing context-aware network, RA-CNN 3, with Xception based LR-CNN and attention based training method achieved superior performance as compared to the BAM based methods.

We also compared our method with a set of context-aware approaches explored in a systemic study on context-aware learning by Sirinukunwattana \textit{et al.} \cite{sirinukunwattana2018improving}. 
Ten different approaches (A-J) were considered to capture contextual information. For a detailed description of these methods please see Figure 2 of \cite{sirinukunwattana2018improving}. It can be observed that these approaches have significant overlap with each other. some of these approaches differ only in term of input patch resolution, as some of these used shared weights for different networks.
First four (A-D) approaches try to capture the context by down-sampling the high-resolution images at different magnification levels e.g. $20\times, 10\times, 5\times, $ and $ 2.5\times$. Channel-wise concatenated four multi-resolution images as an input of a CNN are consider in approach E. However, approaches F and I concatenate the CNN features of four multi-resolution input images in vector form before prediction. The full image at $20\times$ magnification is used as an input in approach H. This approach is not feasible in case of an image with very large spatial dimensions due to memory constraints. Approaches G and J use LSTM to capture the context from the CNN features of four multi-resolution input images. The code of method G is publicly available by the authors of \cite{sirinukunwattana2018improving} and we use that code to retrain the LSTM based method on CRC dataset for a fair comparison. Our method outperformed their context-aware method \cite{sirinukunwattana2018improving} as well for both binary and three-class CRC grading (Table \ref{tab:comp1}).

\begin{figure*}[ht]
   \centering
   \includegraphics[width=0.90\linewidth]{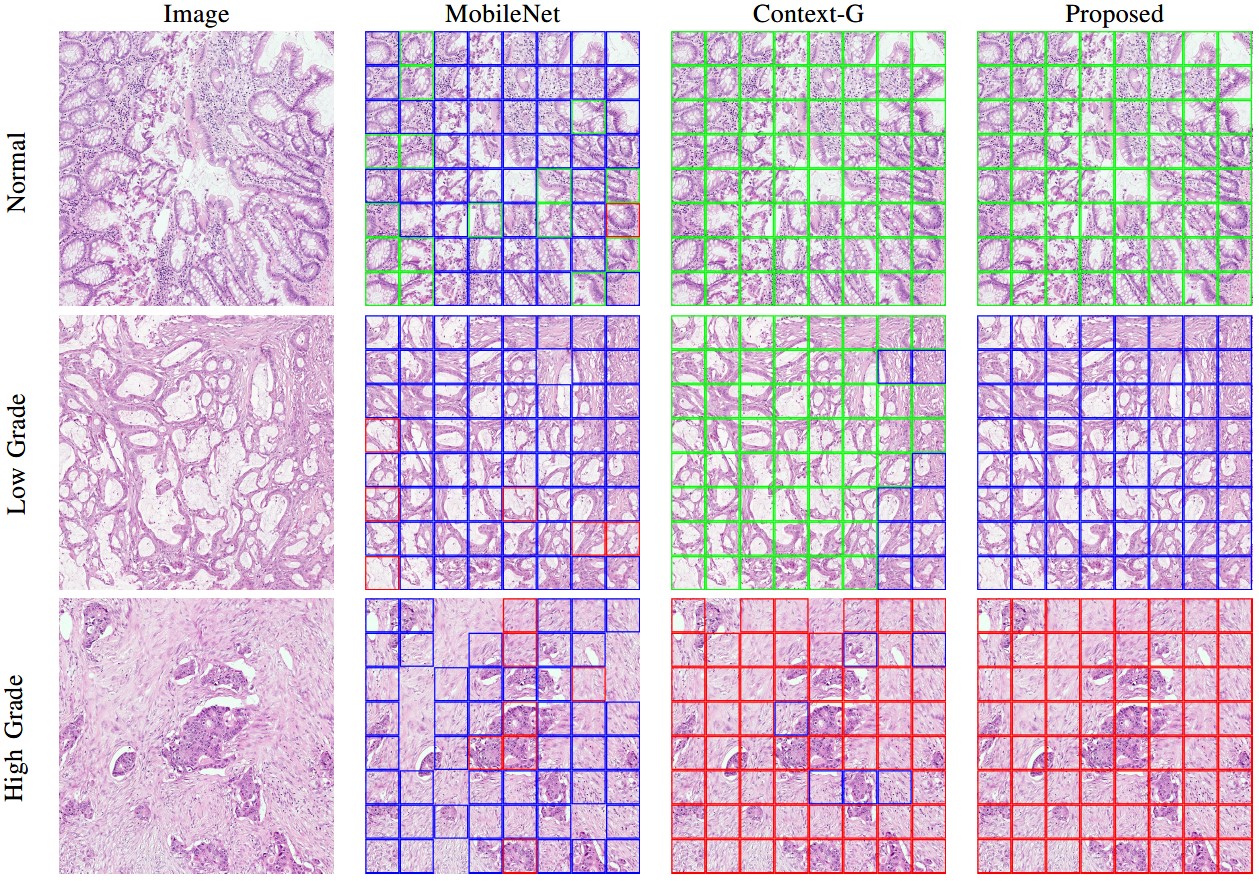}  
    \caption{Visual results of CRC grading are shown for patch classifier, existing context, and the proposed method on an image of size $1,792\times 1,792$.  The stride size for context networks is equal to the size of patch ($224\times 224$) used for patch classifier.  Green, blue and red colors of overlaid rectangular boxes show the normal, low and high-grade predictions respectively, whereas empty box areas represent non-glandular/background regions.}
    \label{fig:vis_res}
\end{figure*}

Visual comparison of best performing patch classifier, Sirinukunwattana \textit{et al.} (Context-G) and the proposed method on three different images with normal, low and high grades are shown in Figure \ref{fig:vis_res}. Patch classifier's prediction is quite irregular for any given image due to the lake of contextual information. The predictions of Context-G are relatively smooth but it predicts the wrong label for the low-grade image which might be due to the use of a low-resolution image for context learning. However, the proposed method predictions are smooth and consistent with the ground truth labels.

We retrained our method on breast cancer dataset to make a direct comparison to all the context-aware approaches presented in \cite{sirinukunwattana2018improving}. Sirinukunwattana \textit{et al.} used a customized network as feature extractor which contains 5 convolution layers with $4\times4$ filter and each layer followed by batch-norm and leaky-ReLU activation. To highlight the significance of RA-CNN, we adopted their network for LR-CNN and trained our RA-CNN 3 with \textit{standard} training strategy in end-to-end training manner. Moreover, the same experimental setup (as in \cite{sirinukunwattana2018improving}) has been used for retraining of our method such as data-splits and F1-score based Rank-sum evaluation metric. Table \ref{tab:comp2} presents the comparison of our proposed method with ten other context-aware approaches. Our method outperformed its counterpart methods with significant margin and achieved best rank-sum scored for breast cancer classification task as well. 

\begin{table}[t]
\centering
\caption{Comparison with state-of-the-art on Colorectal Dataset.}
\begin{tabular}{|l|c|c|}
\hline
\rowcolor[HTML]{EFEFEF}
\multicolumn{1}{|c|}{Method} & Binary (\%)  & Three-class (\%)      \\ \hline
BAM - 1 \cite{RA}              & 95.70 - 2.10 & 87.79 - 2.32          \\ \hline
BAM - 2 \cite{RA}             & 97.12 - 1.27 & 90.66 - 2.45          \\ \hline
Context - G  \cite{sirinukunwattana2018improving}           & 96.44 - 3.61 & 89.96 - 3.54          \\ \hline
ResNet50 \cite{resnet}                    & 98.57 - 1.01 & 92.08 - 2.08          \\ \hline
MobileNet \cite{mobilenet}                   & 97.83 - 1.77 & 92.78 - 2.74          \\ \hline
InceptionV3 \cite{inception}                  & 98.57 - 1.01 & 91.37 - 3.55          \\ \hline
Xception  \cite{xception}                   & 98.58 - 2.01 & 92.09 - 0.98          \\ \hline
Proposed                     & \textbf{99.28 - 1.25}   & \textbf{95.70 - 3.04} \\ \hline
\end{tabular}
\label{tab:comp1}
\end{table}

\begin{table*}[]
\centering
\caption{Rank-sum based comparison as measured by the F1-measure on Breast Dataset.}
\begin{tabular}{|l|c|c|c|c|c|c|c|c|c|c|c|c|}
\hline
\rowcolor[HTML]{EFEFEF}
\multicolumn{1}{|c|}{} & \multicolumn{11}{c|}{Methods}                                                                                                                                                                                                                                                                                                                                         \\ \hline
Classes                & A                             & B                             & C                             & D                             & E                             & F                             & G                             & H                             & I     & J                             & Proposed                \\ \hline
Normal                 & 0.501                         & 0.468                         & 0.523                         & 0.513                         & 0.509                         & \cellcolor[HTML]{FFFC9E}0.603 & \cellcolor[HTML]{FD6864}0.573 & 0.252                         & 0.241 & 0.323                         & \cellcolor[HTML]{FF9F33}0.643                \\ \hline
Benign                 & \cellcolor[HTML]{FD6864}0.453 & \cellcolor[HTML]{FFFC9E}0.468 & \cellcolor[HTML]{FFFC9E}0.482 & \cellcolor[HTML]{FD6864}0.444 & 0.410                         & 0.369                         & 0.423                         & \cellcolor[HTML]{33ACFF}0.489 & 0.333 & \cellcolor[HTML]{FD6864}0.437 & \cellcolor[HTML]{FF9F33}0.511                \\ \hline
InSitu                 & 0.468                         & 0.476                         & 0.486                         & 0.533                         & \cellcolor[HTML]{FF9F33}0.615                & \cellcolor[HTML]{9AFF99}0.614 & \cellcolor[HTML]{FFFC9E}0.581 & 0.286                         & 0.311 & 0.452                         & 0.362                         \\ \hline
Invasive               & 0.401                         & 0.477                         & 0.430                         & \cellcolor[HTML]{FFFC9E}0.54  & \cellcolor[HTML]{33ACFF}0.557 & \cellcolor[HTML]{FFFC9E}0.548 & \cellcolor[HTML]{9AFF99}0.576 & \cellcolor[HTML]{FD6864}0.520                         & 0.446 & \cellcolor[HTML]{FF9F33}0.580 & \cellcolor[HTML]{9AFF99}0.576 \\ \hline
Rank-sum               & 23                            & 22                            & 22                            & 20                            & 16                            & 16                            & 17                            & 20                            & 24    & 18                            & \textbf{10}                   \\ \hline
\end{tabular}
\label{tab:comp2}
\end{table*}

\section{Conclusion}
In this paper, we present a novel context-aware deep neural network for cancer grading, which is able to incorporate 64 times larger context than standard CNN based patch classifiers. The proposed network is well-suited for CRC grading task which relies on recognizing abnormalities in glandular structures. These clinically significant structures vary in size and shape that cannot be captured efficiently with standard patch classifiers due to computational and memory constraints. The proposed context-aware network is comprised of two stacked CNNs. The first LR-CNN is used for learning the local representation of the histology image. The learned local representation is then aggregated considering its spatial pattern by RA-CNN. The proposed context-aware model is evaluated for CRC grading and breast cancer classification. A comprehensive analysis of different variations of the proposed model is presented and compared with existing approaches in the same evaluation setting. The qualitative and quantitative results demonstrate that our method outperformed the patch based classification methodologies, the domain-oriented techniques and existing context-based methods. 
This approached is suitable for cancer analysis which requires large contextual information in the histology images. This includes Gleason grading in prostate cancer and tumor growth pattern classification in lung cancer. Moreover, this approach can further be extended to perform downstream analysis at digital whole slide image level for patient survival analysis.


%





\ifCLASSOPTIONcaptionsoff
  \newpage
\fi



\bibliographystyle{IEEEtran}

\end{document}